# Quantum Key Distribution (QKD) and Commodity Security Protocols: Introduction and Integration

Alan Mink, Sheila Frankel and Ray Perlner

National Institute of Standards and Technology (NIST),
100 Bureau Dr., Gaithersburg, MD 20899
amink@nist.gov, sheila.frankel@nist.gov, ray.perlner@nist.gov

## ABSTRACT

*We present an overview of quantum key distribution (QKD), a secure key exchange method based on the quantum laws of physics rather than computational complexity. We also provide an overview of the two most widely used commodity security protocols, IPsec and TLS. Pursuing a key exchange model, we propose how QKD could be integrated into these security applications. For such a QKD integration we propose a support layer that provides a set of common QKD services between the QKD protocol and the security applications.*

## KEYWORDS

*Quantum Key Distribution, Quantum Networks, IPsec, TLS, Network Security Protocols*

## 1. Introduction

Quantum Key Distribution (QKD) [1, 2] is a technology, based on the quantum laws of physics, rather than the assumed computational complexity of mathematical problems, to generate and distribute provably secure cipher keys over unsecured channels. It does this using single photon technology and can detect potential eavesdropping via the quantum bit error rates of the quantum channel. Sending randomly encoded information on single photons produces a shared secret that is a random string and the probabilistic nature of measuring the photon state provides the basis of its security.

A QKD system consists of a quantum channel and a classical channel. The quantum channel is only used to transmit qbits (single photons) and must consist of a transparent optical path (fiber, free-space and optical switches, no routers, amplifiers or copper). It is a lossy and probabilistic channel. The classical channel can be a conventional IP channel (not necessarily optical), but depending on system design it may need to be dedicated and closely tied to the quantum channel for timing requirements. It would not be unusual for the quantum and classical channels to share a common fiber via wavelength division multiplexing (WDM).

A quantum network connects a number of point-to-point QKD systems together so that one can develop shared secrets between users anywhere on that sub-network. A quantum network would be an embedded sub-network within a conventional communication network for the purpose of developing shared secrets, not transporting secure messages. The physical link (e.g., fiber) that carries the classical channel could certainly support general messages, but not within the QKD classical channel.

Commodity security protocols, such as Internet Protocol Security (IPsec) and Transport Layer Security (TLS, often referred to as Secure Sockets Layer or SSL), currently handle the bulk of today's internet encrypted traffic. Although these protocols are standardized, they have no standard application programming interface (API) and a number of different implementations exist.

With the advent of quantum computers, several of the cryptographic constructs underlying the security model of IPsec and TLS will be broken. Breakthroughs in cryptanalysis continue to present a possible threat as well. Quantum-resistant replacements will have to be found for public-key cryptography and for Diffie-Hellman key agreement. Using quantum keying material within these protocols would solve this problem. Alternate approaches to this problem are being considered via quantum resistant public key cryptographic algorithms [3], although promising, all such algorithms are based on unproven computational complexity assumptions, and if these assumptions are shown to be false, the algorithms will be insecure. QKD does not solve the authentication problem and would rely on conventional authentication techniques (public key or pre-shared secret). QKD does offer perfect forward security and therefore past QKD secrets will remain secure, even when the public key algorithm is broken (since, at most, it was only used for QKD authentication). Although quantum computers are not yet a reality, it is necessary to develop alternative technologies well in advance, since past secrets will be at risk once the threat is realized. There are a few commercial QKD systems and active QKD research programs in Europe [4] and Japan [5] as well as a recently formed European QKD standards effort [6]. This level of activity suggests that QKD and the threats to current cryptography are being taken seriously.

## 2. QKD Overview

The BB84 [7] protocol and its variants are the only known provably secure QKD protocols. Other QKD protocols (e.g., differential phase shift keying [8]), although promising, have yet to be proven secure. The BB84 protocol consists of four stages (See Fig 1). The first stage is the transmission of the randomly encoded single photon stream over the quantum channel from Alice (the sender) to Bob (the receiver) to establish the initial raw key. Alice maintains a temporary database of the state of each photon sent. The second stage is sifting, where Bob sends a list of photons detected and their basis, but not their value, back to Alice over the classical channel. Basis refers to how the photons were measured. Photons can be encoded in one of two bases (e.g., horizontal/vertical or diagonal polarization). There is only one photon and it can only be measured once, so only one basis can be applied. If it's measured in the correct basis the value measured will be correct. If it's measured in the wrong basis, the value will be random. Alice retains, from its database, only those entries received by Bob in the correct basis and sends this revised list back to Bob over the classical channel. Bob retains only those entries on this revised list. Alice and Bob now have a list of sifted keys. These lists are of the same length but may have some errors between them. This is the quantum bit error rate and it is an indication of eavesdropping. The third stage is reconciliation to correct these errors. Cascade [9, 10] and its variants are the predominant reconciliation algorithm that exchange parity and error correcting codes to reconcile errors without exposing the key values. This process requires a number of communications between Bob and Alice, over the classical channel, and results in a list smaller than the sifted list. The fourth stage is privacy amplification, which computes a new (smaller) set of bits from the reconciled set of bits using a hashing algorithm and requires no communication between Alice and Bob. Since the reconciled set of bits were random, the resulting privacy amplified set will also be random. Unless the eavesdropper knows all or most of the original bits, she will not be able to compute the new set.

The benefits of QKD are that it can generate and distribute provably secure keys over unsecured channels and that potential eavesdropping can be detected. QKD is not subject to threats from quantum computers or break through algorithms that can defeat the current computationally complex key exchange methods. Because QKD generates random strings for shared secrets, attaining a QKD system and reverse engineering its theory of operation would yield no mechanism to defeat QKD. QKD can use existing optical media infrastructure for both quantum and classical channels, but the quantum channel photons cannot pass through amplifiers or routers. Optically transparent switches are okay and thus switched networks of QKD systems are possible and have been demonstrated.

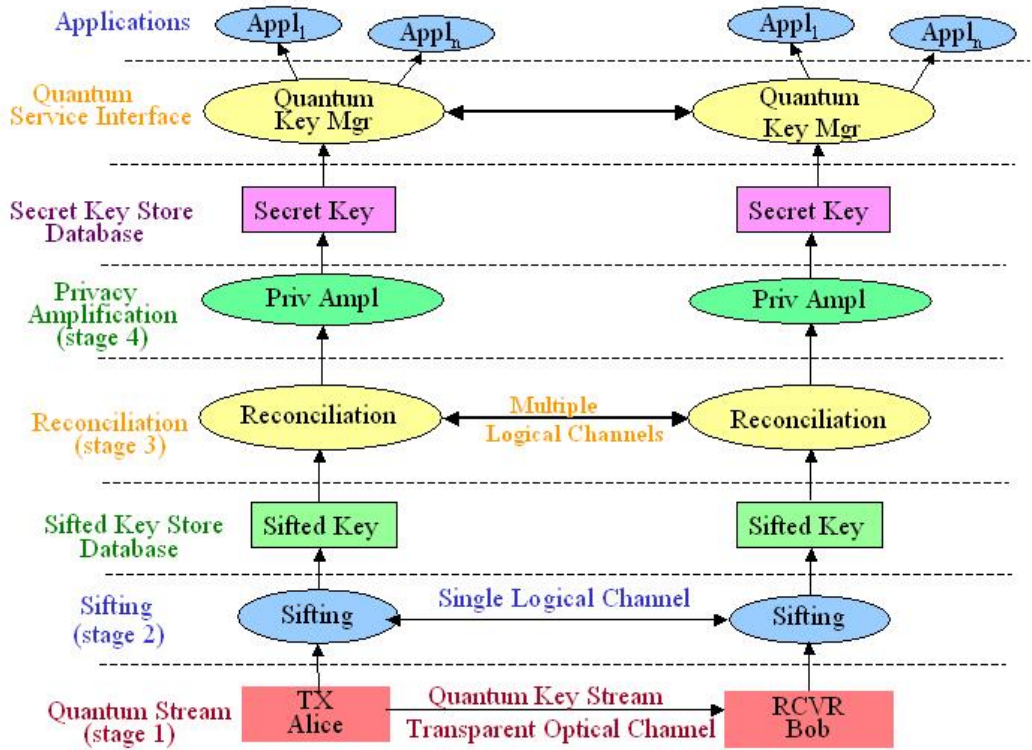

Figure 1. QKD protocol flow

The limitations of QKD are that it's currently a costly technology and requires dedicated hardware, it's still in its infancy, it's restricted in distance until quantum repeaters are developed, its key generation rate decreases with distance, engineered realization of components are subject to attack and it requires an optically transparent path for the quantum channel. The end-to-end distance can be extended by concatenating multiple short distance QKD systems to span a longer distance, however this does require a multi-hop propagation of the key to the distant end point or sequential encryption/decryption of the message along this path. In either case, the intermediate points must all be trusted systems for this strategy to be successful.

QKD requires its own sub-network to generate keys, although this sub-network may use/share existing communication infrastructure (i.e., existing fiber and transparent optical switches). Unlike current key exchange methods that are purely software based and have the range of the entire network, QKD is limited to the sub-network that connects its hardware. So if QKD were invoked between two nodes in the network, some additional checking would be needed to verify that both nodes can support QKD and that there is a QKD sub-network connection between them that is operational.

Although security proofs exist for the theoretical BB84 protocol, no engineering models exist for it. Such engineering models would be necessary to determine when an implementation's parameters stray too far from the theoretic and are no longer covered by the security proofs. Real systems based on engineered components that are not identical and do not operate exactly as the theory requires have been shown to be vulnerable to attack [11]. Without solutions to these limitations, QKD may not be a viable alternative security technology and may be limited to niche markets. Engineering shrinks and mass production could make the hardware small and more affordable, while future research may solve many of the current limitations.

A small QKD network [12] has been demonstrated that interfaced with IPsec, but did not include the use of a One-Time Pad cipher [13]. Another small QKD network [14, 15] using a One-Time Pad cipher was demonstrated, but it did not use any existing security protocol utility (e.g., IPsec, TLS, etc). The encryption/decryption was built directly into the application. What is needed as these QKD networks emerge, grow and eventually interact, are guidelines and standards as to how they should integrate with existing network infrastructure and security protocols.

## 3. IPsec Overview

IPsec [16, 17] is a suite of protocols that provides security to Internet communications at the network layer, and thus IPsec must also handle any non-malicious errors in the data stream. The most common current use of IPsec is to provide a Virtual Private Network (VPN), either between two locations (gateway-to-gateway) or between a remote user and an enterprise network (host-to-gateway). IPsec can also provide end-to-end security.

+ Security controls exist for network communications at various layers of the ISO model. Controls applied at the network layer apply to all applications and are not application-specific. For example, all network communications between two hosts or networks can be protected at this layer without modifying any applications on the clients or the servers. Network layer controls also provide a way for network administrators to enforce certain security policies. Another advantage of network layer controls is that since IP information (e.g., IP addresses) is added at this layer, the controls can protect both the data within the packets and the IP information for each packet. However, network layer controls provide less control and flexibility for protecting specific applications than transport and application layer controls.

IPsec is a protocol with the following components:

+ Two security protocols, Authentication Header (AH) [18] and Encapsulating Security Payload (ESP) [19] comprise the base IPsec protocol. The IPsec ESP header provides confidentiality and/or integrity protection. The AH header provides integrity protection without confidentiality. Both ESP and AH provide data origin authentication, access control, and, optionally, replay protection and/or traffic analysis protection. ESP is used much more frequently than AH because of its encryption capabilities, as well as other operational advantages. For a VPN, which requires confidential communications, ESP is the natural choice.

+ Internet Key Exchange (IKE) protocol is IPsec's preferred key management protocol. IKE [20, 21] negotiates the cryptographic algorithms and related settings to be used for AH and ESP. IPsec also uses IKE to negotiate IPsec connection settings; authenticate endpoints; define the security parameters of IPsec-protected connections; and manage, update, and delete IPsec-protected communication channels.

The protections provided by IPsec, and the protection that IKE provides to its own key exchange traffic, require the use of cryptographic algorithms, which include encryption algorithms, MACs (Message Authentication Codes, used for integrity protection), and prfs (Pseudo-Random Functions, which generate secret keys and other values used within the IPsec protocols). IPsec security mechanisms are not tied to any specific cryptographic algorithms. Standard default algorithms are, however, specified in order to support interoperability [22, 23].

In order to use automated key management protocols such as IKE to negotiate and manage IPsec protections and secret keys between two peers, those peers must be able to definitively authenticate each other (i.e. verify each others' identities) in the course of the IKE negotiation. The methods used within IKE are: pre-shared secret keys, Public Key Certificates, or, for host-to-gateway IPsec, a combination of a certificate for the gateway and an Extensible Authentication Protocol (EAP)-based authentication method for the host.

An IKE negotiation generally consists of two round-trip exchanges between the peers, referred to as the initiator and the responder. In the first exchange, the initiator proposes a list of cryptographic algorithms acceptable to the initiator for the protection of further IKE traffic (a protected channel called the IKE Security Association, or IKE SA), and sends its own public Diffie-Hellman (DH) value. The responder selects a single set of cryptographic algorithms from that list, and sends its public DH value. At this point, the peers are ready to generate the shared secret that will be used to generate all of the secret keys. Once a shared secret is established a two-step process is used to generate the set of secret keys:

    SKEYSEED = prf(randoms, DH shared secret)

    Secret Keys = prf(SKEYSEED, randoms)

where randoms include nonces and other information. Additional iterative calculations may be required to supply the requisite amount of keying material. The keys are used to protect the IKE SA and the IPsec-protected data between the peers (i.e. the IPsec or child SA). Once these keys are generated, all further IKE traffic is encrypted and integrity-protected. In the second exchange, the peers authenticate each other's identities by signing or MACing some data: all of the previous messages exchanged by the peers, plus the peers' nonces (random values) and other values. The peers also negotiate the set of cryptographic algorithms that will be used to provide IPsec protection to data exchanges between the peers, and the secret keys that will provide this protection are calculated.

## 4. TLS Overview

The Transport Layer Security (TLS) protocol [24] is based on the earlier Secure Sessions Layer (SSL) protocol and is explicitly invoked by an application. Although it is now used in a wide variety of applications, its most common use is for the encryption of traffic between a web server and a browser. It is assumed that a mechanism such as the transmission control protocol (TCP) is in place to correct any non-malicious errors in the data stream.

TLS is made up of two phases:

- **The TLS handshake protocol** allows two parties to agree upon a key agreement method, as well as the encryption and MAC algorithms that will be used to protect the data, collectively known as a ciphersuite. The two parties then use the key agreement method to agree upon a shared master secret -- one or both of the parties may be authenticated during this process. Finally, the master secret is used to derive encryption and MAC keys. The TLS record protocol uses the established keys to encrypt and integrity protect the data packets along with a sequence number (to prevent malicious reordering of the packets in the data stream.)

- **The TLS record protocol** handles TLS I/O, data compression, encryption/decryption, and MAC and key generation.

TLS security mechanisms are not tied to any specific cryptographic algorithms and implement both public key and pre-shared secret key trust models. Standard default algorithms are, however, specified in order to support interoperability. The first round trip exchange of TLS handshake protocol establishes the ciphersuite to be used during a TLS session and subsequent exchanges are used to establish a shared secret and authenticate one or both of the two parties.

Once a shared secret (referred to in TLS as the Pre-master Secret) is established a two-step process is used to generate the set of secret keys:

>    Master Secret = prf (Pre-Master Secret, randoms, nonces)

>    Secret Keys = prf(Master Secret, randoms, additional nonces)

where randoms are random values chosen by each of the parties during the first round trip of the handshake and nonces are non-random strings specified by the protocol to prevent identical keys from being chosen in different contexts. Then the protocol proceeds directly to the TLS record protocol to send encrypted and integrity protected data.

## 5. QKD Service Interface

In order to function properly, any system using QKD needs to transport both quantum and classical data from a specified source to a specified destination, resolve competing requests for shared hardware, and manage shared keys between neighboring trusted nodes via a multi-hop mechanism. For classical data, similar functionality can be provided by standard lower level protocols such as TCP/IP and Ethernet, however the unique demands of transporting quantum data intact require us to design a new service interface in order to provide the same functionality to the quantum portion of the protocol flow. This service interface should provide the following services: a Network Layer protocol to manage the QKD sub-network, QKD Key Synchronization/Demultiplexing and QKD multi-hop keying. As with other QKD stages, messages between service layer peers will need to be integrity-protected, while network layer messages for routing configuration may not. We can view the QKD protocol flow as an independent application that produces a database of secure keys at both ends of a link. The service interface distributes synchronized key from the QKD secure key database to the security applications. The QKD protocols may use conventional security applications (e.g., IPsec or TLS) or apply its own for authentication and integrity-protection of its classical messages.

Network management features of a QKD service interface include handling switching, routing, and QKD protocol startup operations along with normal operations of the quantum devices. When a QKD connection between two nodes is requested by an application, this utility should determine (via routing tables) if such a connection is possible (are there QKD links and are they operational) and, if switching is necessary, can it be accomplished (depending on current use and priorities). Complications arise when switching (vs. static fixed links) is required, because QKD currently uses circuit switching that requires 10s of seconds to switch, initialize and produce usable keys.

Besides managing the network and quantum devices, the QKD service interface should provide Synchronization and Demultiplexing of the quantum key stream. Demultiplexing divides up the bits in the QKD key store into multiple independent key streams to each application. Synchronization makes sure that the same bits, in the same order, are allocated to the same demultiplexed stream on both sides of the network connection. This also requires a mechanism to detect and recover from loss of synchronization.

QKD is a point-to-point protocol on a single link. If a quantum network topology results in communication end points that aren't directly connected, such as a mesh network where there are multiple connected QKD links, then a Multi-Hop mechanism is necessary to deliver the QKD key across the network to the end point node. For example, see Fig 2, if QKD links A-B, C-D and E-F are three QKD links in a quantum network that connect node 1, 2, 3 and 4, and B and C are co-located in the same node as are D and E. If we want to send a message from node 1 to node 4, encrypted with key(a), then we need to get key(a) to node 4, but it exists only on Nodes 1 and 2. So node 2 gets key(c) and One-Time Pad encrypts key(a) with key(c) and then sends that encrypted key(AC) as a message to node 3. Node 3 decrypts it using key(c), extracting the original key(a). Node 3 then gets key(e) and One-Time Pad encrypts key(a) with

key(e) and then sends that encrypted key(ae) as a message to node 4. Node 4 decrypts it using key(e), extracting the original key(a). This multi-hop mechanism relies on the security of the intermediate nodes (2 and 3) and may not always be acceptable for some applications. So whether two nodes are directly connected or require multi-hop key transport, is information necessary to the application so it can decide if the QKD connection is acceptable.

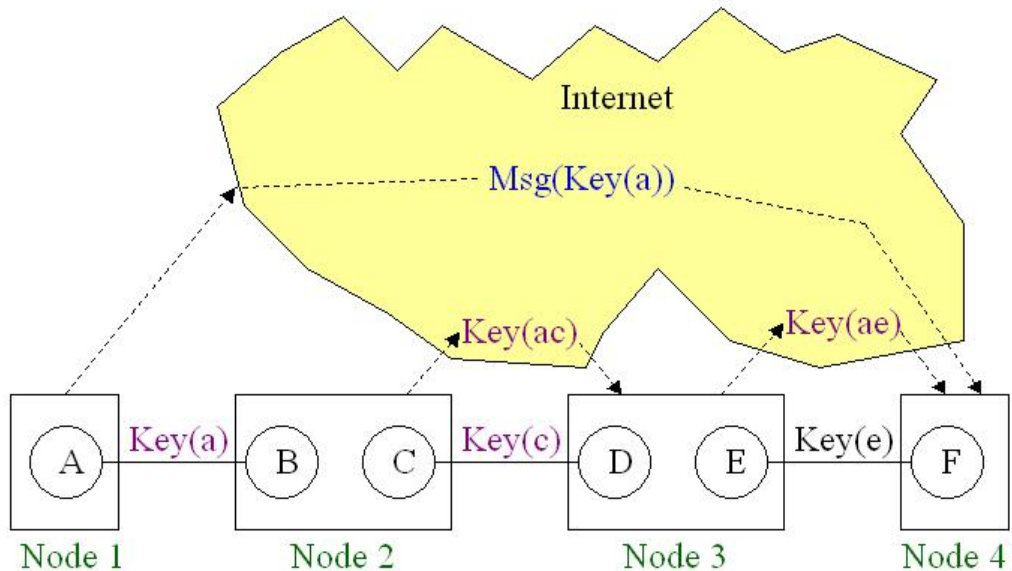

Figure 2. Example of QKD multi-hop mechanism across 3 QKD links

## 6. Integration of QKD Material into IPsec/TLS

There are two models we can apply for interfacing QKD with security applications. One is where the QKD service interface provides keys to the application and lets the application handle encryption, authentication and transport of the message, as they currently do. Thus the QKD service interface acts as a key exchange function to the security applications. The other model is to keep the keys within the QKD service interface and require the application to pass it the message to be encrypted, authenticated and transported end-to-end. This would require the security application to specify to the QKD service interface the cipher for encryption and the MAC for integrity algorithms, as well as the end point destination of the message. Although either model should be able to attain the same level of security, we believe that letting the QKD support layer process the messages duplicates the existing functionality of the security protocols, possibly introducing unintended consequences, and bypasses all of the specifications and validation efforts that they have demonstrated.

Pursuing the key exchange model, we attempt to minimize the effect to the existing protocol by constraining the QKD integration to the smallest part of the application possible. One approach to integrating QKD with IPsec is to keep QKD within IKE rather than spawning a parallel key exchange function. IKE would access the QKD key when directed, rather than calculate the key. This approach serves two purposes. First, it avoids having to develop another key exchange function, fraught with duplication of effort, bugs and unintended consequences. Second it keeps QKD encapsulated and external from the security protocol itself. Although TLS doesn't have an explicit key exchange function like IKE, that function is embedded in the protocol and QKD integration should minimize any impact.

**The use of QKD Material in IPsec and TLS**

Both IPsec and TLS develop a shared secret and then use that to compute keys for encryption and integrity protection. Thus, QKD Material could be used within TLS, IPsec and IKE as the shared secret, the keys or as a One-Time Pad:

A. A Quantum Key could be used to establish the shared secret for a TLS session or an IKE SA. In this scenario, quantum keys would replace the DH shared secret that is calculated in IKE and either the premaster secret or the master secret in TLS, thus eliminating the need to calculate a DH shared secret.

B. Quantum Keys could be used to protect the IKE SA's traffic and/or the peers' IPsec-protected data or to replace TLS encryption and MAC keys. Instead of calculating these keys from the shared secret, Quantum Keys could be used for both encryption and integrity-protection. These keys would be used in conjunction with the negotiated encryption and integrity-protection algorithms, thus eliminating the need to both calculate a shared secret and compute secret keys.

C. As an alternative to option B, Quantum Keying Material could be used as a One-Time Pad, requiring an amount of key equal to the message size for each type of protection that is provided. A One-Time Pad could be used to protect the IKE SA's traffic and/or the peers' IPsec-protected data as well as the TLS bulk encryption process. In this case, the One-Time Pad would be used to perform encryption instead of a conventional encryption algorithm. For integrity-protection, the One-Time Pad would have to be used in conjunction with a standard MAC integrity-protection algorithm, such as UMAC.

Since symmetric key algorithms are quantum-resistant, option A, together with a quantum-resistant authentication method (pre-shared secret keys, a Kerberos-based method, or the use of Public Key Certificates that use quantum-resistant algorithms) would enable IKE or TLS to negotiate traditional symmetric keys in a quantum-resistant manner. Options B or C would add an extra dimension of security: keys that would be resistant to discovery at the point when current symmetric-key encryption algorithms will no longer be strong enough to resist attacks. This would provide protection to data that requires extremely long-term resistance to discovery.

**Changes to IKE and the TLS handshake protocol**

In the course of an IKE or TLS handshake ciphersuite negotiation, it is necessary to agree on the selection of several cryptographic mechanisms. These include:

- The encryption algorithm(s) (e.g., AES-CBC) that will protect data exchanges

- The MAC(s) (e.g., HMAC-SHA) that will provide integrity protection to traffic

- The PRF (e.g., HMAC-SHA for IPsec – TLS 1.1 specifies a standard PRF) that will be used to generate the secret keys

- The key exchange method (e.g., a 2048-bit MODP DH group for IPsec, or a TLS ciphersuite specifying the use of RSA in TLS) that will be used to generate the shared secret

For cases A, B and C above, IKE and TLS would either have to negotiate using a new value for an existing cryptographic attribute (e.g., One-Time Pad as an encryption algorithm) or a new cryptographic attribute (e.g., use of Quantum Keys for encryption) in addition to the regular set of attributes.

Before this could be done, IKE and TLS would need to verify that a functioning quantum channel was available between the peers. If the answer were negative, IKE and TLS would not propose any of these quantum attributes. Since failures and stalls can occur during a security session, exchanges must incorporate timeout procedures. IKE exchanges include standardized procedures for handling situations in which expected responses are not received within a pre-defined time, but TLS requests and exchanges do not and will need to include them.

The negotiated PRF is used by IKE and TLS to iteratively derive all of the keys required by the IKE SA, the IPsec SA and TLS. If some or all of those keys are replaced by quantum keying material or by a One-Time Pad, IKE and TLS would need to adjust the key length calculations accordingly. If only quantum keys and One-Time Pads will be used as keys, and the peers are not willing to negotiate the use of conventional keys for any purpose, IKE still needs to negotiate a PRF, since the PRF is also used for peer authentication within IKE.

Similarly, the DH shared secret is used as input to the key calculations. If no key calculations are necessary, it is not necessary to negotiate the DH group attribute for the IKE SA or TLS.

The randomly generated nonces (IKE) and randoms (TLS) that are used in the conventional key calculations are part of TLS and IKE's security infrastructure. They ensure liveness of negotiations, protecting against replay attacks. Furthermore, each participant must sign the peer's nonce as part of the peer authentication process. Thus, even if all of the keys derive from quantum key material, the peers must still generate and send nonces and randoms.

### Changes to IPsec and the TLS record protocol

Once the secret keys are calculated by IKE, they are passed over to IPsec for secure storage within the IPsec SA Database. Similarly TLS stores its keys in its Database. Using quantum keying material in IKE or TLS, instead of calculated keys would thus not require any changes to their normal processing. Similarly, IPsec's SA Database or TLS's Database would not need to have any indication that the source of the keys differed from the normal source, since the keys would be used in conjunction with their conventional cryptographic algorithms.

The use of a One-Time Pad is a different matter. Normally, the secret keys are stored in IPsec's SA Database. When it is time to require a new key, IPsec notifies IKE, and a new SA, with fresh keying material, is negotiated. For the One-Time Pad, it would not be practical for IKE to initially supply IPsec with sufficient keying material to last for the whole lifetime of the SA. IKE would not necessarily know how much keying material would be required. In addition, sufficient keying material might not be available when the SA is initially negotiated. Finally, storage space for the totality of keying material could be a problem. A more reasonable solution would be for the IPsec routines to periodically request fresh keying material. This would necessitate some sort of communication mechanism between the quantum key generation mechanism and IPsec.

TLS supports a stream cipher that is similar to a One-Time Pad. It uses the cipher key stored in the TLS database to compute a key stream via a repeated computation similar to that of computing the key from the shared secret. Using QKD keys for this purpose would require a stream interface to QKD keys rather than this computational function. The encryption process would become an interactive process rather than a deterministic algorithm that simply derives a keystream from a fixed length key.

### Possible Complications

When IKE, IPsec or TLS requests quantum keying material, it would be problematic if sufficient key material is not available. For cases A and B above, an IKE or TLS negotiation could fail if it times out while waiting for the delivery of initial keying material. This would necessitate another negotiation, which might succeed if more keying material has become available. But it could also cause repeated failures, resulting in blocking the traffic that needs

IPsec or TLS protection. The peers would then have to choose whether to fall back to conventional cryptographic protection, or to postpone the communications until sufficient quantum keying material is available. If sufficient quantum keying material is not available during negotiations, and the peers are not willing to use conventional cryptographic mechanisms instead, the negotiation will fail, and a security association will not be established.

Use of One-Time Pads, case C above, for IPsec and TLS keys presents a different problem. In this case, after a TLS session or IPsec security association is established, an undetermined and potentially large amount of key may be required. What if, at some point, keying material for the One-Time Pad is requested, but not available? Either the peers would have to delay the data communications, or they would then have to choose whether to fall back to conventional cryptographic protection, or to postpone the communications until sufficient quantum keying material is available. This presents a new situation to these protocols, since currently once the initial security association or session is established all keys are algorithmically generated internally. Renegotiating at this point would require additions to the protocol.

In both cases, sane heuristics for re-sending and/or re-negotiating would need to be set in place.

**Protecting the BB84 exchange**

IPsec or TLS, with conventional keys, can be used to integrity protect the BB84 exchange. Each IPsec SA is defined with specific traffic selectors that determine to which types of traffic the SA will be applied. For example, a single IPsec SA could protect all traffic between designated IP addresses and/or ports. By using a set of standard ports for the QKD classical channels IPsec can protect that traffic. Since TLS is an application specific protocol, a separate instance of TLS would need to be programmed into the QKD protocol flow.

Using IPsec or TLS with QKD keys to integrity protect the BB84 exchange using quantum keys presents a bootstrapping problem. Since quantum key material will not be available until the completion of a number of BB84 exchanges, TLS and IPsec must use conventional keys until a sufficient amount of quantum key material is developed. Manually established keys could be installed, or conventional keys could be generated via a standard IKE or TLS negotiation. However, manual keys do not scale well as the network expands.

The BB84 exchange does not require confidentiality protection; its only security requirement is integrity-protection, to prevent data tampering. Thus, one would not need to negotiate an encryption algorithm. Also, the amount of key needed for integrity-protection is small so it should not use much of the quantum key supply.

## 7. Summary

We have presented overviews of the QKD protocol and the widespread internet security applications, IPsec and TLS. Pursuing a key exchange model, we proposed how QKD could be integrated into these security applications. We also note that existing security protocols could be used to authenticate and integrity protect QKD protocol messages, but care must be taken to avoid the use of quantum keys before they exist. Finally we discussed a QKD service interface between the QKD protocol and the security applications. This interface provides a set of common QKD services needed by existing security protocols. QKD may not develop into a viable widely deployable technology, but with on-going research and commercially available systems, it's too early to dismiss it.

## ACKNOWLEDGEMENTS

The identification of any commercial product or trade name does not imply endorsement or recommendation by the National Institute of Standards and Technology

## REFERENCES

[1] Gisin N, Ribordy G, Tittel W and Zbinden H 2002 Quantum cryptography *Rev. Mod. Phys*, **4** 41.1-41.8

[2] Chou C W, Laurat J, Deng H, Choi K S, de Riedmatten H, Felinto D and Kimble H J 2007 Functional quantum nodes for entanglement distribution over scalable quantum networks *Science* **316** 1316-20

[3] R. Perlner and D. Cooper, "Quantum Resistant Public Key Cryptography: A Survey", Proc of IDtrust 2009, Gaithersburg, MD, Apr. 14-19, 2009.

[4] SECQO, "QKD Network Demonstration and Conference", Vienna, Austria, Oct 8-10, 2008.

[5] Proc. Updating Quantum Cryptography 2008 Conf., Tokyo, Japan, Dec 1-2, 2008 <http://www.rcis.aist.go.jp/events/uqc2008/>

[6] European Telecommunications Standards Institute (ETSI), Standards work started in Dec 2008 < http://etsi.org/WebSite/homepage.aspx> (select "Committees & Portals", then select "ISG" and then select "QKD")

[7] C. H. Bennet and G. Brassard, "Quantum cryptography: Public key distribution and coin tossing," in Proc IEEE Intern'l Conf on Computers, Systems and Signal Processing, Bangalore, India, 1984, pp. 175-179.

[8] K. Inoue, E. Woks and Y. Yamamoto, "Differential phase shift quantum key distribution" Phys. Rev. Lett. 89, 037902, (2002).

[9] Gilles Brassard and Louis Salvail, "Secret-Key Reconciliation by Public Discussion*", proceedings of Eurocrypt'94, Lecture notes in computer Science*, 765, Spriger Verlag, 410-423.

[10] A. Nakassis, J. Bienfang, and C. Williams, "Expeditious reconciliation for practical quantum key distribution," Proce of SPIE Quantum Information and Computation II, Volume 5436, Orlando Florida, 12-14 Apr 2004, pp 28-35.

[11] V. Scarani and C. Kurtsiefer, "The black paper of quantum cryptography: real implementation problems", arXiv:0906.4547v1 [quant-ph], June 2009, <http://arxiv.org/abs/0906.4547>

[12] C. Elliott, A. Colvin, D Pearson, O. Pikalo, J. Schlafer and H. Yeh, "Current Status of the DARPA Quantum Network", March 2005, <http://arxiv.org/ftp/quant-ph/papers/0503/0503058.pdf>.

[13] http://en.wikipedia.org/wiki/One-time_pad

[14] L. Ma, A. Mink, H. Xu, O. Slattery and X. Tang, "Experimental Demonstration of an Active Quantum Key Distribution Network with Over Gbps Clock Synchronization", IEEE Communication Letters, Vol 11(12): 1019~1021 (2007)

[15] A.. Mink, L. Ma, H. Xu, O. Slattery, B. Hershman and X. Tang, "A Quantum Network Manager That Supports A One-Time Pad Stream", Proc of the 2[nd] Intern'l Conf on Quantum, Nano, and Micro Technology, St. Luce, Martinique, Feb 10-15, 2008, pp 16-21.

[16] S. Kent and K. Seo, "Security Architecture for the Internet Protocol", RFC 4301, Dec. 2005.

[17] S. Frankel, "Demystifying the IPsec Puzzle", Artech House, 2001, 273 pp.

[18] S. Kent, "IP Authentication Header", RFC 4302, Dec. 2005.

[19] S. Kent, "IP Encapsulating Security Payload (ESP)", RFC 4303, Dec. 2005.

[20] C. Kaufman, ed., "Internet Key Exchange (IKEv2) Protocol", RFC 4306, Dec. 2005.

[21] P. Eronen and P. Hoffman, "IKEv2 Clarifications and Implementation Guidelines", RFC 4718, Oct. 2006.

[22] V. Manral, "Cryptographic Algorithm Implementation Requirements for Encapsulating Security Payload (ESP) and Authentication Header (AH)", RFC 4835, Apr. 2007.

[23]     J. Schiller, "Cryptographic Algorithms for Use in the Internet Key Exchange Version 2 (IKEv2)", RFC 4307, Dec. 2005.

[24]     T. Dierks and E. Rescorla, "The Transport Layer Security (TLS) Protocol, Version 1.2", RFC 5246, Aug. 2008.

**Authors**

**Alan Mink** is an electronic engineer in the Advanced Networking Division at The National Institute of Standards and Technology. He holds a PhD in electrical engineering and has been a lecturer at the University of Maryland, where he taught computer architecture. He has worked on projects in the areas of computer aided design, networking, parallel processing, performance measurement, cluster computing, digital TV and quantum communications.

**Sheila Frankel** is a computer scientist in the Computer Security Division at the National Institute of Standards and Technology. She holds a Masters degree in computer science. She participates in the Internet Engineering Task Force (IETF) IPsec standardization effort, and was responsible for NIST's IPsec/IKE reference implementation and interactive Web-based interoperability tester. She is the author of a book on IPsec, "Demystifying the IPsec Puzzle". Currently, she is involved with the Federal Government's transition to IPv6, the next generation Internet protocol.

**Ray Perlner** is a Mathematician in the Computer Security Division at The National Institute of Standards and Technology. He holds bachelor's degrees in Physics and Mathematics. He has worked on projects in electronic authentication, cryptographic protocols, and cryptanalysis.